\documentclass[lettersize,journal]{IEEEtran}
\usepackage{amsmath,amsfonts}
\usepackage{array}
\usepackage{textcomp}
\usepackage{stfloats}
\usepackage{url}
\usepackage{verbatim}
\usepackage{graphicx}
\usepackage{cite}
\usepackage{booktabs}
\usepackage{multirow}
\usepackage{makecell}
\usepackage{subfigure}
\usepackage[misc]{ifsym} 
\usepackage{balance}
\usepackage{listings}
\usepackage{color, xcolor}
\usepackage{algorithm2e}
\usepackage{diagbox}
\usepackage{tcolorbox}
\usepackage{stfloats}
\usepackage{enumitem}
\usepackage{cleveref}
\usepackage{ulem}
\usepackage{colortbl} 

\definecolor{lightcoral}{rgb}{0.94, 0.5, 0.5}
\definecolor{lightgreen}{rgb}{0.56, 0.93, 0.56}
\definecolor{harvestgold}{rgb}{0.85, 0.57, 0.0}
\definecolor{brightlavender}{rgb}{0.75, 0.58, 0.89}
\definecolor{capri}{rgb}{0.0, 0.75, 1.0}
\definecolor{carminepink}{rgb}{0.92, 0.3, 0.26}
\definecolor{celadon}{rgb}{0.67, 0.88, 0.69}
\definecolor{darkpastelgreen}{rgb}{0.01, 0.75, 0.24}
\definecolor{DeepSkyBlue4}{RGB}{0,104,139}

\definecolor{acccolor}{RGB}{238, 245, 252} % 极淡的矢车菊蓝
\definecolor{latcolor}{RGB}{242, 242, 242} % 极淡的灰

\begin{document}

\title{From Feedback Loops to Policy Updates: Reinforcement Fine-Tuning for \\ LLM-Based Alpha Factor Discovery}

\author{
	\IEEEauthorblockN{Lingzhe Zhang, Tong Jia\IEEEauthorrefmark{1}, Yunpeng Zhai, Zixuan Xie, Chiming Duan,\\ Minghua He, Philip S. Yu,~\IEEEmembership{Fellow,~IEEE} and Ying Li\IEEEauthorrefmark{1},~\IEEEmembership{Member,~IEEE}}
	\thanks{Lingzhe Zhang, Tong Jia, Chiming Duan, Minghua He and Ying Li are with Peking University, Beijing, China. Yunpeng Zhai is with Alibaba Group, China. Zixuan Xie are with Nanjing University, Nanjing, China. Philip S. Yu are with University of Illinois Chicago, United States.}
	\thanks{Email: \{zhang.lingzhe, duanchiming, hemh2120\}@stu.pku.edu.cn, \{jia.tong, li.ying\}@pku.edu.cn, zhaiyunpeng.zyp@alibaba-inc.com and psyu@uic.edu}
	\thanks{* Corresponding author: Tong Jia, e-mail: (jia.tong@pku.edu.cn); Ying Li, e-mail: (li.ying@pku.edu.cn)}
}
\maketitle

\begin{abstract}
	Modern quantitative trading increasingly relies on systematic models to extract predictive signals from large-scale financial data, where alpha factor discovery plays a central role in transforming market observations into tradable signals. Recent LLM-based methods have shown promise in automating factor generation, but most of them still rely on prompt-level generation--evaluation--feedback loops for iterative optimization. As the loop becomes longer, repeatedly appended historical candidates and feedback can cause context explosion, increase inference cost, dilute useful information, and introduce feedback drift. Moreover, these methods often depend on very large LLMs whose stable generation preferences may lead to structurally similar expressions, redundant candidates, and search stagnation. To address these limitations, we propose \textsc{QuantEvolver}, a self-evolving alpha factor discovery framework based on reinforcement fine-tuning. Instead of accumulating feedback in the prompt, \textsc{QuantEvolver} converts executable quantitative evaluation into policy updates, enabling a Miner LLM to internalize historical optimization experience through parameter learning. Specifically, \textsc{QuantEvolver} constructs high-quality seed factors, builds diverse seed--time-window training tasks, generates executable Factor DSL expressions, evaluates them through Regime Backtest, and optimizes the Miner LLM with Diversity-Complementarity Reward. During training, high-quality factors are continuously accumulated in a Mined Factor Database, which serves as the final discovered factor library. Extensive experiments on three realistic market benchmarks demonstrate the effectiveness of \textsc{QuantEvolver}, which consistently improves the primary evaluation metric of each task over existing LLM-based alpha factor discovery baselines, produces higher-quality and more complementary factor pools.
\end{abstract}


\section{Introduction}

Modern quantitative trading increasingly relies on systematic models to extract predictive signals from large-scale financial data. Unlike many conventional prediction settings, financial markets are highly noisy, weakly structured, and non-stationary: useful signals are often sparse, unstable across regimes, and easily overwhelmed by spurious correlations. As a result, the central challenge in quantitative research is not merely to fit historical observations, but to continuously discover robust and economically meaningful signals that can generalize across assets, time periods, and market conditions.

Alpha factors provide a fundamental abstraction for this challenge. An alpha factor is a numerical signal designed to capture predictive patterns related to future returns, risks, or cross-sectional asset rankings. However, alpha factor discovery remains a challenging search problem. First, the candidate space is combinatorially large, as factors can be composed from diverse market variables, mathematical operators, temporal windows, and normalization schemes. Second, financial signals are typically weak, noisy, and regime-dependent, making it difficult to distinguish genuinely predictive factors from spurious historical correlations. Third, useful factors should not only achieve strong empirical performance, but also remain robust across time periods and market conditions, while providing non-redundant information beyond existing signals. These challenges make the scalable and reliable discovery of alpha factors a central problem in quantitative finance.

To address the difficulty of alpha factor discovery, extensive studies have explored automated factor mining beyond manual factor engineering. Existing non-LLM methods generally fall into two categories. The first category adopts rule-based symbolic search, such as genetic programming, symbolic regression, and operator-template enumeration~\cite{zhang2020autoalpha, ren2024alpha, kakushadze2016101, arnaldo2014multiple, yang2026alpha, ren2024riskminer}. These methods represent alpha factors as symbolic expressions and explore the candidate space through predefined operators, mutation rules, crossover operations, or syntactic constraints. The second category introduces learning-guided generation, where trainable models or reinforcement learning policies are used to construct, rank, or select factor expressions based on historical evaluations or previously discovered factors~\cite{yu2023generating, xu2024text, zhu2025alphaqcm, shi2025alphaforge, chen2025alphasage, yang2026alpha}. Although these methods reduce manual engineering effort, their search processes are still largely shaped by predefined operators, action spaces, expression templates, or task-specific objectives. As a result, they often struggle to flexibly compose factor expressions from high-level market semantics, scenario descriptions, and prior research hypotheses.

Large language models (LLMs) offer a promising way to alleviate these limitations. With their capabilities in symbolic reasoning, code-like generation, and instruction following~\cite{zhang2025survey, zhang2025log, zhang2024towards, zhang2024multivariate, zhang2024reducing, zhang2025scalalog, zhang2025agentfm, zhang2025thinkfl, zhang2026agentic, zhang2025logdb, zhang2025xraglog, zhang2025surveyparallel, zhang2024time, kang2022separation, zhang2025adaptive, liu2025ora, zhang2025microremed, pan2025omni, he2025walk, pan2025d, hong2025cslparser, he2025united, zhang2026hypothesize, huang2025uda, liu2025aaad, duan2025logaction, zhang2026runtimeslicer, zhang2026efficient, zhang2026e2e, xiao2025coorlog, zhang2026towards}, LLMs can generate and refine alpha expressions from alpha libraries, market contexts, human instructions, and empirical feedback. Existing LLM-based frameworks have therefore evolved from interactive factor generation to feedback-driven and search-augmented alpha mining. Early systems, such as Alpha-GPT~\cite{wang2025alpha} and FAMA~\cite{li2024can}, use LLMs as factor miners or interactive assistants. Later frameworks, such as QuantAgent~\cite{wang2024quantagent}, AlphaBench~\cite{luoalphabench}, and AlphaAgent~\cite{tang2025alphaagent}, further involve LLMs in factor evaluation and refinement by incorporating historical experiment records, human knowledge, or market insights. More recent search-augmented and agentic methods, including LLM-powered MCTS~\cite{shi2026navigating}, R\&D-Agent~\cite{li2026r}, and QuantaAlpha~\cite{han2026quantaalpha}, combine LLM generation with backtesting feedback, multi-agent workflows, or evolutionary exploration. 

\begin{figure}[htbp]
	\centering
	\includegraphics[width=1\linewidth]{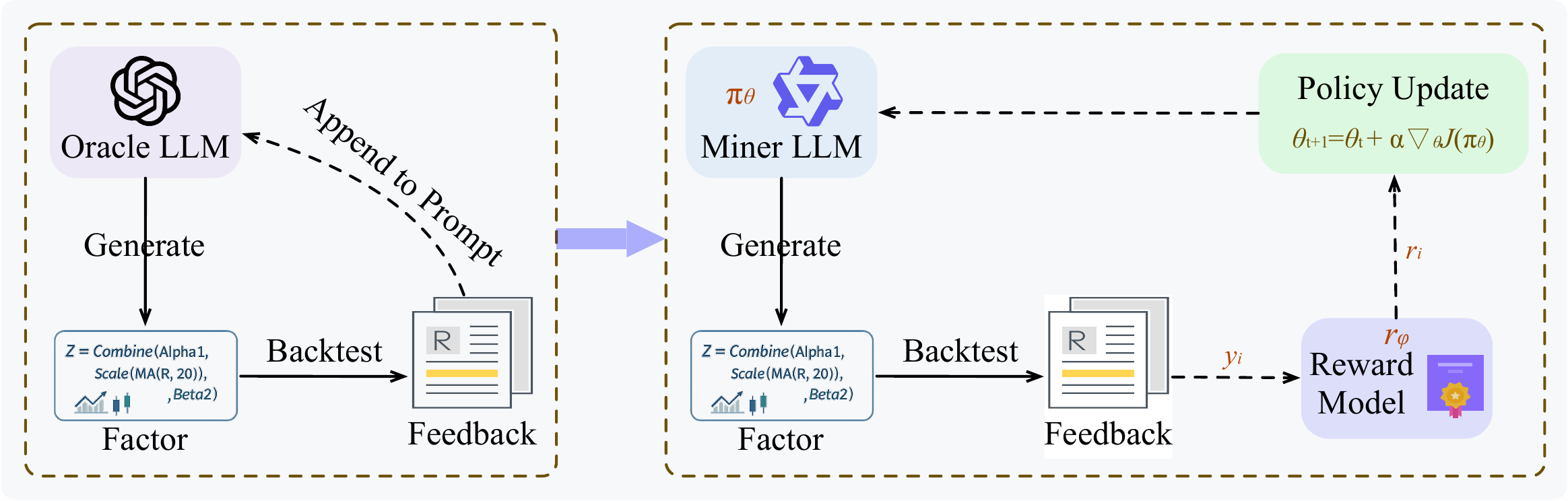}
	\caption{LLM-Based Alpha Factor Discovery: From Feedback Loops to Policy Updates.}
	\label{fig: intro-example}
\end{figure}

Although these LLM-based methods have demonstrated promising results in generating effective alpha factors, they still largely rely on prompt-level feedback loops for iterative optimization, as illustrated in Figure~\ref{fig: intro-example}. This design is flexible and easy to implement, but becomes increasingly inefficient and unstable when scaled to large factor mining campaigns, leading to two key challenges:

\begin{itemize}[leftmargin=*]
	\item \textbf{Context explosion and feedback drift.}
	Existing LLM-based factor mining methods often rely on multi-round generation--evaluation--feedback loops, where historical candidates, evaluation results, and feedback signals are repeatedly appended to the prompt. As the loop becomes longer, the accumulated context increases inference cost, dilutes useful feedback, and introduces irrelevant or outdated information. This makes the feedback-driven optimization process increasingly unstable and may gradually drift away from the original factor mining objective.
	
	\item \textbf{Large-model dependence and search stagnation.}
	Existing methods typically depend on very large LLMs to generate syntactically valid and semantically meaningful factor expressions. However, such models often exhibit stable generation preferences inherited from pretraining and prompting, repeatedly producing structurally similar expressions or reusing familiar operator patterns. This limits exploration diversity, causes redundant factor candidates, and weakens the ability to discover genuinely novel alpha signals at scale.
\end{itemize}

To fill these gaps, we propose \textsc{QuantEvolver}\footnote{https://github.com/QuantLLM/QuantEvolver}, a self-evolving alpha factor discovery framework based on reinforcement fine-tuning.

To address \textbf{Challenge 1}, \textsc{QuantEvolver} replaces prompt-level feedback accumulation with reinforcement fine-tuning. As illustrated in Figure~\ref{fig: intro-example}, instead of appending long histories of generated factors, evaluation results, and refinement feedback to the prompt, each training round only uses bounded generation--evaluation feedback. The optimization experience accumulated across rounds is then internalized into the policy model through parameter updates. In this way, \textsc{QuantEvolver} uses implicit optimization memory to replace explicit long-context feedback histories, reducing context growth and mitigating feedback drift during large-scale factor mining.

To address \textbf{Challenge 2}, \textsc{QuantEvolver} adopts a lightweight LLM (parameter $\le$ 30B) as the base policy and continuously evolves its generation distribution through executable quantitative rewards. Each generated DSL factor expression is compiled and evaluated on historical market data, and the resulting reward is used to update the policy toward more effective factor proposals. Moreover, we design \textit{DiCo Reward}, a \textit{Di}versity-\textit{Co}mplementarity reward mechanism that calibrates raw predictive performance with novelty, redundancy, behavioral diversity, and complementarity signals. This reward design encourages the policy to generate factors that are not only predictive, but also structurally diverse, behaviorally distinct, and complementary to existing candidates.

Beyond these two designs, \textsc{QuantEvolver} further introduces a seeded task construction strategy to stabilize reinforcement fine-tuning. Specifically, an oracle LLM is first used to construct a high-quality seed pool by generating candidate factor expressions, checking DSL validity, empirically scoring candidates, and selecting elite non-redundant seeds. The selected seeds are then expanded by a task bank builder into regime-aware training tasks, where each task contains a seed expression, time split, factor family, mutation hints, and scenario metadata. This design provides meaningful starting points and structured training contexts, enabling the lightweight LLM to focus on improving and diversifying promising factor candidates rather than searching from scratch.

We conduct experiments on three realistic market benchmarks spanning short-horizon single-asset direction prediction, high-frequency cross-sectional factor mining, and daily equity ETF factor discovery. The results show that \textsc{QuantEvolver} consistently improves the primary evaluation metric of each task over existing LLM-based alpha factor discovery methods. Specifically, \textsc{QuantEvolver} achieves a 7.8\% relative improvement in directional accuracy on the single-asset direction prediction task. On the high-frequency cross-sectional benchmark, it exceeds the strongest baseline by 109.5\% in best out-of-sample RankIC and by 186.9\% in top-10 out-of-sample RankIC mean. On the daily equity ETF benchmark, \textsc{QuantEvolver} further produces a high-quality and fuseable factor pool, achieving strong out-of-sample RankIC under a conventional equity-market evaluation protocol. These results demonstrate that \textsc{QuantEvolver} is effective across heterogeneous factor discovery settings, from high-frequency digital-asset prediction to daily equity factor mining. In summary, the key contributions of this work are as follows:

\begin{itemize}[leftmargin=*]
	\item We introduce a policy-update-driven paradigm for LLM-based alpha factor discovery. Specifically, we replace prompt-level feedback accumulation with reinforcement fine-tuning, allowing a small language model to internalize historical optimization experience through parameter updates and progressively improve its factor generation policy.
	
	\item Built upon the above paradigm, we propose \textsc{QuantEvolver}, a self-evolving alpha factor discovery framework. \textsc{QuantEvolver} combines oracle-based seed construction, regime-aware task bank building, executable quantitative evaluation, and DiCo Reward to provide stable training tasks and reward signals that encourage predictive, diverse, and complementary alpha factors.
	
	\item We validate the effectiveness of \textsc{QuantEvolver} on three realistic market benchmarks. Extensive experiments show that \textsc{QuantEvolver} surpasses state-of-the-art LLM-based methods on the primary evaluation metrics, especially in terms of factor quality, diversity, and out-of-sample effectiveness.
\end{itemize}

\section{Background}

In this section, we present the essential background of this paper, including the formal definition of alpha factor discovery and the emerging use of reinforcement fine-tuning (RFT) for adapting LLMs to task-specific feedback.

\subsection{Alpha Factor Discovery}

Alpha factors are numerical signals designed to capture predictive patterns in financial markets. Given historical market observations, such as prices, volumes, returns, and derived technical indicators, an alpha factor maps these observations into a scalar signal that can be used for forecasting future returns, ranking assets, or constructing trading portfolios. At a fundamental level, an alpha factor can be viewed as a function from market observations to numerical signals. Therefore, many alpha factors can be naturally expressed in a formulaic form, where market variables are composed through mathematical operators, time-series transformations, and cross-sectional operations.

For example, a simple momentum-style factor may compute the recent return trend of an asset, while a reversal-style factor may rank assets according to short-term price declines. A formulaic alpha factor can be expressed as a symbolic expression, such as Equation~\ref{eq:example_factor}.

\begin{equation}
	f = \mathrm{rank}(\mathrm{ts\_mean}(\mathrm{return}, 5)),
	\label{eq:example_factor}
\end{equation}

Here, $\mathrm{ts\_mean}$ computes a moving average over a five-period window and $\mathrm{rank}$ normalizes the resulting values across assets. Such expressions are compact and executable: once the variables and operators are defined, the expression can be directly applied to historical market data to produce factor values.

Formally, let $\mathcal{X}$ denote historical market data and let $\mathcal{L}$ denote a domain-specific language (DSL) that defines the valid variables, operators, and composition rules for constructing alpha expressions. An alpha factor can be represented as an expression $f \in \mathcal{L}$, which maps market data $\mathcal{X}$ to a factor value sequence as Equation~\ref{eq:definition}.

\begin{equation}
	z_f = f(\mathcal{X}).
	\label{eq:definition}
\end{equation}

Here, the output $z_f$ denotes the factor values produced by applying expression $f$ to market data $\mathcal{X}$. Given a target market outcome $y$, such as future return or cross-sectional asset ranking, the empirical quality of $f$ can be measured by an evaluation function $M(z_f, y)$. Common evaluation signals include predictive accuracy, information coefficient (IC), rank information coefficient (RankIC), and portfolio-level performance indicators.

In practical quantitative pipelines, alpha factors are rarely used in isolation. Instead, multiple factors are usually combined into a factor set and further fused into a composite signal for prediction, ranking, or portfolio construction. Given a discovered factor set as shown in Equation~\ref{eq:factor_set}.

\begin{equation}
	\mathcal{F} = \{f_1, f_2, \ldots, f_n\}, \quad f_i \in \mathcal{L},
	\label{eq:factor_set}
\end{equation}

The corresponding factor values can be denoted as Equation~\ref{eq:factor_matrix}.

\begin{equation}
	\mathbf{Z}_{\mathcal{F}} = [z_{f_1}, z_{f_2}, \ldots, z_{f_n}],
	\label{eq:factor_matrix}
\end{equation}

Here, each $z_{f_i}$ is computed according to Equation~\ref{eq:definition}. A downstream fusion model or aggregation function $g$ can then combine these factor values into a final predictive signal as Equation~\ref{eq:fusion}.

\begin{equation}
	s = g(\mathbf{Z}_{\mathcal{F}}).
	\label{eq:fusion}
\end{equation}

It illustrates why factor discovery is not only about finding individually strong factors. If newly discovered factors are highly redundant, they may provide limited incremental value after fusion. Therefore, an effective factor set should contain factors that are predictive, diverse, and complementary to each other.

The goal of alpha factor discovery is to search for a set of factor expressions $\mathcal{F}$ that achieves strong empirical performance under validation protocols and provides useful signals for downstream fusion. This problem is challenging because the candidate space grows rapidly with the number of variables, operators, temporal windows, and composition patterns. Moreover, financial signals are noisy and regime-dependent, so factors that appear effective in one period may fail to generalize in another. In practice, discovered factors are also expected to provide incremental value beyond existing factor libraries, since highly similar expressions often contribute limited new information.

\subsection{Reinforcement Fine-Tuning}

Reinforcement fine-tuning (RFT) is a training paradigm that adapts language models using task-specific reward signals. Unlike supervised fine-tuning (SFT), which learns from reference outputs or demonstrations, RFT allows a model to generate candidate outputs, receive rewards from an external evaluator, and update its policy according to the evaluated quality of these outputs. This makes RFT suitable for structured generation tasks where high-quality outputs are difficult to exhaustively annotate but can be assessed through task-specific feedback, such as mathematical reasoning, code generation, and tool-use planning~\cite{christiano2017deep, ziegler2019fine}.

Formally, given an input prompt $x$, a language model policy $\pi_\theta$ generates an output $y \sim \pi_\theta(\cdot \mid x)$. A reward function $R(x,y)$ evaluates the quality of the generated output. The objective of RFT is to update the policy parameters $\theta$ to maximize the expected reward as Equation~\ref{eq:rft_objective}.

\begin{equation}
	\max_{\theta} \;
	\mathbb{E}_{x \sim \mathcal{D},\, y \sim \pi_\theta(\cdot \mid x)}
	\left[ R(x,y) \right].
	\label{eq:rft_objective}
\end{equation}

Here, RFT converts task feedback into parameter-level policy updates. Through repeated optimization, the model can gradually internalize task-specific preferences and improve its generation behavior, rather than relying only on in-context examples or manually written feedback prompts.

Modern RFT methods differ in how rewards are collected and how policy updates are performed. Proximal Policy Optimization (PPO)~\cite{schulman2017proximal} is widely used because it stabilizes policy updates through clipped objectives or trust-region-style constraints. Direct Preference Optimization (DPO)~\cite{rafailov2023direct} simplifies preference optimization by directly learning from preference pairs without explicitly training a separate reward model. More recently, Group Relative Policy Optimization (GRPO)~\cite{shao2024deepseekmath} has been used to improve reasoning-oriented language models by comparing multiple sampled completions under the same prompt. Specifically, for a prompt $x$, GRPO samples a group of $K$ outputs as Equation~\ref{eq:grpo_group}.

\begin{equation}
	\mathcal{Y}(x) = \{y_1, y_2, \ldots, y_K\}, \quad
	y_i \sim \pi_{\theta_{\mathrm{old}}}(\cdot \mid x).
	\label{eq:grpo_group}
\end{equation}

Each output receives a reward $R(x,y_i)$ from a task-specific evaluator. The relative advantage of each output can then be estimated by normalizing rewards within the group as shown in Equation~\ref{eq:grpo_advantage}.

\begin{equation}
	\hat{A}_i =
	\frac{R(x,y_i) - \mathrm{mean}(\{R(x,y_j)\}_{j=1}^{K})}
	{\mathrm{std}(\{R(x,y_j)\}_{j=1}^{K})}.
	\label{eq:grpo_advantage}
\end{equation}

Here, GRPO encourages the model to increase the probability of outputs that outperform other candidates generated under the same prompt, while suppressing relatively poor outputs. This group-wise comparison is useful for structured generation tasks, where multiple candidates may be syntactically valid but differ substantially in task-level quality.

Despite the success of RFT in reasoning, code generation, and other structured generation tasks, its use in LLM-based alpha factor discovery remains underexplored. Existing LLM-based alpha mining methods mainly rely on prompt-level interaction and iterative feedback, while few studies investigate whether reinforcement fine-tuning can directly improve the factor generation policy itself.

\begin{figure*}[htbp]
	\centering
	\includegraphics[width=1\linewidth]{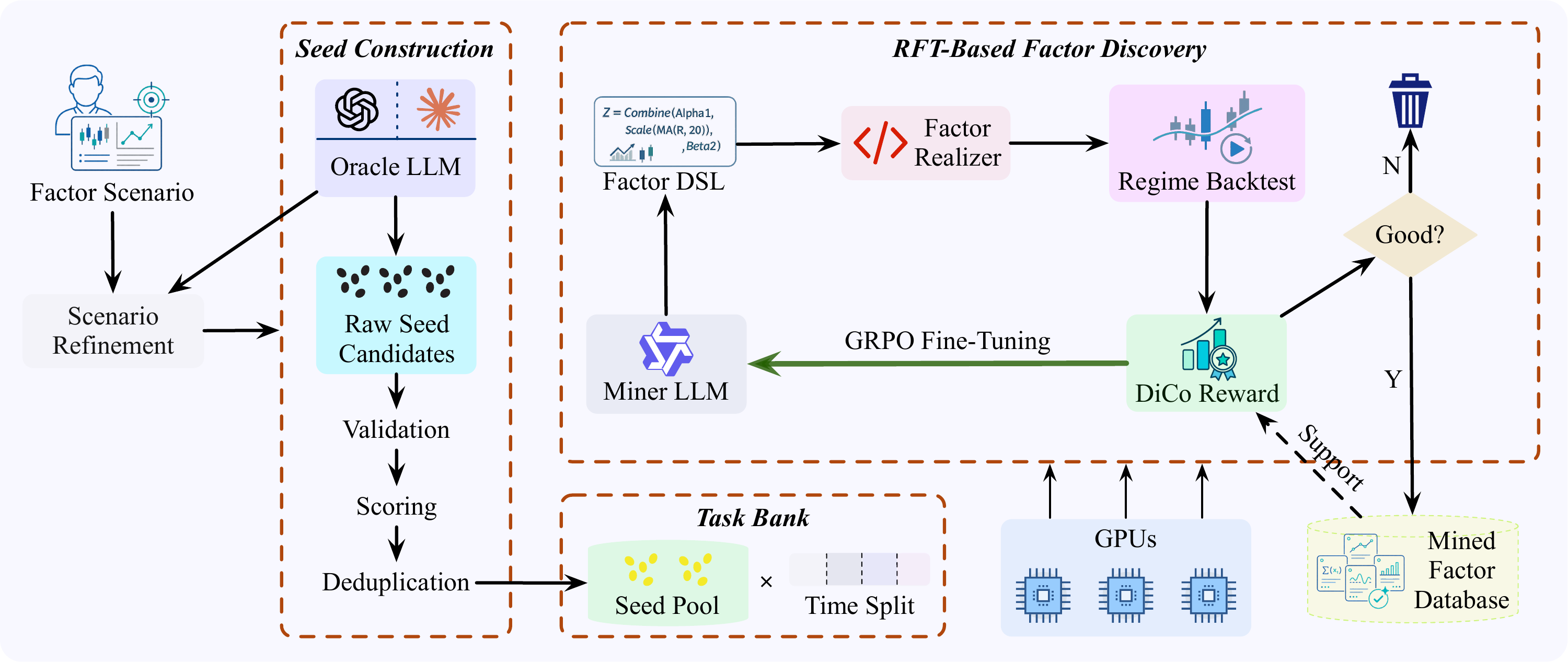}
	\caption{Overall Framework of \textsc{QuantEvolver}.}
	\label{fig:pipeline}
\end{figure*}

\section{Methodology}

The workflow of \textsc{QuantEvolver} is illustrated in Figure~\ref{fig:pipeline}. Given a user-specified factor mining scenario, \textsc{QuantEvolver} first performs seed construction, where the raw scenario is refined into a structured factor scenario and an oracle LLM generates raw seed candidates. These candidates are then validated, scored, and deduplicated to form a high-quality seed pool. By combining the selected seeds with different time splits in the backtesting period, \textsc{QuantEvolver} further constructs a task bank that provides diverse and structured training instances for subsequent reinforcement fine-tuning.

Based on the task bank, \textsc{QuantEvolver} conducts RFT-based factor discovery with a miner LLM. The miner LLM iteratively generates candidate factor expressions, receives executable quantitative feedback, and is updated through in-the-loop GRPO fine-tuning. To guide this process, \textsc{QuantEvolver} introduces DiCo Reward (Diversity-Complementarity Reward), which encourages the model to discover factors with strong predictive quality, diversity, and complementarity. High-quality factors generated throughout training are continuously stored in a mined factor database, and the final output is a validated factor library tailored to the target scenario.

\subsection{Seed Construction}

The goal of seed construction is to provide high-quality starting points for reinforcement fine-tuning. Instead of asking the miner LLM to explore the entire factor expression space from scratch, \textsc{QuantEvolver} constructs a compact seed pool from a user-specified factor mining scenario or a prepared scenario configuration. This design improves the stability of subsequent policy optimization and guides the miner LLM toward promising regions of the factor space.

Given a raw user scenario $u$, \textsc{QuantEvolver} may first perform scenario refinement to convert it into a structured factor scenario as Equation~\ref{eq:scenario_refinement}, where $s$ contains task-specific information such as the target market, prediction horizon, available variables, factor constraints, and evaluation objective.

\begin{equation}
	s = \phi_{\mathrm{refine}}(u),
	\label{eq:scenario_refinement}
\end{equation}

Based on the structured scenario $s$, a strong LLM or a prepared seed source proposes a set of raw seed candidates as Equation~\ref{eq:raw_seed_generation}, where each $c_i$ is a candidate factor expression written in the predefined factor expression language.

\begin{equation}
	\mathcal{C}_{\mathrm{raw}}
	=
	\{c_1, c_2, \ldots, c_m\}
	=
	\phi_{\mathrm{seed}}(s),
	\label{eq:raw_seed_generation}
\end{equation}

The raw candidates are then filtered through DSL validation, empirical scoring, and canonical deduplication. First, DSL validation checks whether each candidate expression satisfies the syntax, operator, and type constraints of the factor language. Let $\mathcal{L}$ denote the valid expression space. The valid candidate set is defined as Equation~\ref{eq:seed_validation}.

\begin{equation}
	\mathcal{C}_{\mathrm{valid}}
	=
	\{c_i \in \mathcal{C}_{\mathrm{raw}} \mid c_i \in \mathcal{L}\}.
	\label{eq:seed_validation}
\end{equation}

Second, each valid candidate is executed on historical market data and assigned an empirical score. Given market data $\mathcal{X}$ and an evaluation metric $M(\cdot)$, the score of candidate $c_i$ is computed as Equation~\ref{eq:seed_scoring}.

\begin{equation}
	q(c_i) = M(c_i(\mathcal{X})).
	\label{eq:seed_scoring}
\end{equation}

Here, $q(c_i)$ measures the initial empirical quality of the candidate under the target scenario. The specific metric is selected according to the benchmark objective, such as DirAcc for directional prediction or RankIC for cross-sectional ranking.

Third, \textsc{QuantEvolver} removes redundant candidates to avoid constructing a seed pool dominated by duplicated or near-identical expressions. In implementation, candidates are ranked by their empirical scores and greedily selected using canonical expression signatures, such as normalized AST hashes. A candidate is retained only if it passes the quality filter and its canonical form has not already been selected as Equation~\ref{eq:seed_pool}, where $\tau_q$ is the quality threshold, $h(\cdot)$ denotes the canonical expression signature, and $\mathcal{H}_{\mathrm{selected}}$ is the set of signatures of already selected seeds. The resulting seed pool $\mathcal{S}$ contains valid, effective, and non-redundant factor expressions.

\begin{equation}
	\mathcal{S}
	=
	\mathrm{TopK}
	\left(
	\{c_i \in \mathcal{C}_{\mathrm{valid}}
	\mid
	q(c_i) \geq \tau_q,
	\;
	h(c_i) \notin \mathcal{H}_{\mathrm{selected}}
	\}
	\right)
	\label{eq:seed_pool}
\end{equation}

After obtaining the seed pool, \textsc{QuantEvolver} constructs a task bank for reinforcement fine-tuning. Let Equation~\ref{eq:time_windows} denote a set of evaluation windows in the backtesting period.

\begin{equation}
	\mathcal{W}
	=
	\{w_j\}_{j=1}^{n}
	=
	\{(\mathrm{start}_j,\mathrm{end}_j)\}_{j=1}^{n}
	\label{eq:time_windows}
\end{equation}

Each window specifies the market interval on which generated factors are evaluated during RFT. For each seed $f \in \mathcal{S}$ and each time window $w_j \in \mathcal{W}$, \textsc{QuantEvolver} constructs a seeded factor mining task as Equation~\ref{eq:seeded_task}, where $f$ is the seed expression, $s$ is the structured scenario, $w_j$ is the evaluation window, and $o$ denotes the corresponding factor discovery objective. 

\begin{equation}
	\tau_{f,j}
	=
	\left(
	f,\;
	s,\;
	w_j,\;
	o
	\right)
	\label{eq:seeded_task}
\end{equation}

The full task bank is then defined as the Cartesian product between the seed pool and the evaluation windows as Equation~\ref{eq:task_bank}.

\begin{equation}
	\mathcal{B}
	=
	\{\tau_{f,j} \mid f \in \mathcal{S},\; w_j \in \mathcal{W}\}.
	\label{eq:task_bank}
\end{equation}

Each task in $\mathcal{B}$ provides a structured training instance for the miner LLM. It specifies which seed expression should be improved, under which market scenario it should be optimized, and on which historical window it should be evaluated. By combining seeds with multiple evaluation windows, the task bank exposes the miner LLM to diverse market conditions and reduces the risk that reinforcement fine-tuning overfits to a single historical interval. This seeded task construction allows \textsc{QuantEvolver} to focus policy optimization on improving and diversifying promising factor candidates, rather than performing unconstrained exploration from scratch.

\subsection{RFT-Based Factor Discovery}

After constructing the task bank, \textsc{QuantEvolver} performs RFT-based factor discovery to optimize the factor generation distribution of the \emph{Miner LLM}. Different from prompt-level feedback loops that repeatedly append previous candidates and feedback into the context, \textsc{QuantEvolver} uses reinforcement fine-tuning to convert executable quantitative evaluation into parameter updates. As illustrated in Figure~\ref{fig:pipeline}, the Miner LLM generates candidate expressions in the \emph{Factor DSL}, the \emph{Factor Realizer} converts them into executable factor computations, the \emph{Regime Backtest} module evaluates their empirical performance on historical market windows, and high-quality factors are accumulated in the \emph{Mined Factor Database}.

Formally, let $\mathcal{B}$ denote the task bank constructed in the previous subsection. Each task $\tau \in \mathcal{B}$ contains a seed expression, a structured market scenario, an evaluation window, and a factor discovery objective. Given a task $\tau$, the Miner LLM, modeled as a policy $\pi_\theta$, generates a candidate factor expression as Equation~\ref{eq:policy_generation}.

\begin{equation}
	\hat{f} \sim \pi_\theta(\cdot \mid \tau).
	\label{eq:policy_generation}
\end{equation}

Here, $\hat{f}$ denotes a newly generated or mutated factor expression conditioned on the task information. In the main setting, $\hat{f}$ is required to follow the predefined Factor DSL syntax. For DSL-free ablations, the same framework can instead evaluate generated Python Backtrader factor code.

To make the generated candidate evaluable, \textsc{QuantEvolver} uses the Factor Realizer to parse and convert the Factor DSL expression into executable factor code. Let $\psi_{\mathrm{realize}}(\cdot)$ denote the realization function. The executable implementation of $\hat{f}$ is obtained as Equation~\ref{eq:factor_realizer}, where $p_{\hat{f}}$ is executable code that can be applied to historical market data. If the generated output violates the required format, fails Factor DSL validation, contains unsupported operators, or cannot be executed safely, it is treated as invalid and assigned a low reward.

\begin{equation}
	p_{\hat{f}} = \psi_{\mathrm{realize}}(\hat{f}),
	\label{eq:factor_realizer}
\end{equation}

For each valid executable candidate $p_{\hat{f}}$, \textsc{QuantEvolver} performs Regime Backtest on the historical evaluation window specified by $\tau$. Let $\mathcal{X}^{\tau}$ denote the market data in this window and let $y^{\tau}$ denote the corresponding future-return target. The factor values generated by $p_{\hat{f}}$ are computed as Equation~\ref{eq:factor_execution}.

\begin{equation}
	z_{\hat{f}}
	=
	p_{\hat{f}}\left(\mathcal{X}^{\tau}\right).
	\label{eq:factor_execution}
\end{equation}

The Regime Backtest module then compares the generated factor values against the target outcome as Equation~\ref{eq:regime_backtest}, where $\psi_{\mathrm{bt}}(\cdot)$ denotes the task-specific backtesting function and $e_{\hat{f}}$ contains empirical evaluation results, such as DirAcc, IC, RankIC, coverage, validity status, and auxiliary statistics. The primary metric depends on the benchmark objective: for example, DirAcc is used for single-asset directional prediction, while RankIC is used for cross-sectional ranking or single-asset rank correlation tasks.

\begin{equation}
	e_{\hat{f}}
	=
	\psi_{\mathrm{bt}}
	\left(
	z_{\hat{f}},
	y^{\tau},
	\tau
	\right),
	\label{eq:regime_backtest}
\end{equation}

Based on the backtesting result, \textsc{QuantEvolver} computes the training reward as Equation~\ref{eq:dico_interface}, where $\mathcal{A}$ denotes the current Mined Factor Database and auxiliary archives used during training. The detailed design of $R_{\mathrm{DiCo}}$ is introduced in the next subsection.

\begin{equation}
	r_{\hat{f}}
	=
	R_{\mathrm{DiCo}}
	\left(
	\hat{f},
	e_{\hat{f}},
	\tau,
	\mathcal{A}
	\right),
	\label{eq:dico_interface}
\end{equation}

During training, candidates that pass empirical-quality and coverage requirements are inserted into the Mined Factor Database. Let $\eta(\hat{f}, e_{\hat{f}})$ be a selection function that determines whether a generated candidate should be retained as Equation~\ref{eq:mined_archive_update}.

\begin{equation}
	\mathcal{A}
	\leftarrow
	\mathcal{A}
	\cup
	\left\{
	\hat{f}
	\mid
	\eta(\hat{f}, e_{\hat{f}})=1
	\right\}.
	\label{eq:mined_archive_update}
\end{equation}

In implementation, this database-level selection checks whether the candidate is executable, satisfies minimum metric and coverage thresholds, and is not an exact duplicate of an already saved factor. Candidates that do not pass the selection criterion are discarded after their rewards have contributed to policy optimization.

The Miner LLM is updated using group-based reinforcement fine-tuning with GRPO. For each task $\tau$, the Miner LLM samples a group of $K$ candidate outputs as Equation~\ref{eq:factor_group}.

\begin{equation}
	\mathcal{G}_{\tau}
	=
	\{
	\hat{f}_1, \hat{f}_2, \ldots, \hat{f}_K
	\},
	\quad
	\hat{f}_k \sim \pi_\theta(\cdot \mid \tau).
	\label{eq:factor_group}
\end{equation}

Each candidate $\hat{f}_k$ is realized by the Factor Realizer, evaluated by Regime Backtest, and assigned a reward $r_k$ according to Equation~\ref{eq:dico_interface}. These rewards are then passed to the GRPO optimizer, which performs group-wise policy updates as described in the background section. This process increases the likelihood of candidates that receive higher executable rewards under the same task, while suppressing invalid or low-quality generations.

Overall, RFT-based factor discovery treats reinforcement fine-tuning as a search-distribution optimization mechanism. The Miner LLM is continuously improved through policy updates, but the final output of \textsc{QuantEvolver} is not merely the fine-tuned model. Instead, high-quality factors discovered throughout training are accumulated in the Mined Factor Database, and the final result is a validated factor set tailored to the target scenario.

\subsection{DiCo Reward}

DiCo Reward, short for Diversity-Complementarity Reward, is designed to guide RFT-based factor discovery beyond pure predictive performance. A reward based only on empirical score may cause the Miner LLM to repeatedly exploit a small set of similar expressions. Therefore, DiCo Reward augments the task-specific predictive reward with lightweight shaping terms that encourage structural diversity and behavioral complementarity. The required expression-level, family-level, and behavior-level records are maintained in the Mined Factor Database.

Given the generated factor $\hat{f}$ and its Regime Backtest result $e_{\hat{f}}$, DiCo Reward first extracts a task-specific predictive reward $r_{\mathrm{pred}}(\hat{f}) = g(e_{\hat{f}})$, where $g(\cdot)$ extracts the primary metric from the backtesting result. Depending on the benchmark, this metric can be directional accuracy, IC, RankIC, or mean cross-sectional RankIC. Invalid, non-executable, or insufficient-coverage candidates receive a low reward.

To discourage direct repetition, DiCo Reward applies an exact-repeat penalty $r_{\mathrm{exact}}(\hat{f}) = -\lambda_{\mathrm{exact}} \mathbb{I}_{\mathrm{exact}}(\hat{f})$, where $\mathbb{I}_{\mathrm{exact}}(\hat{f})$ indicates whether the normalized expression hash of $\hat{f}$ has already appeared in the Mined Factor Database. This prevents the policy from receiving the same reward for repeatedly generating the same expression.

Beyond exact duplicates, many factors may share the same structural template while differing only in local constants, such as lookback windows. \textsc{QuantEvolver} therefore maintains structural family records in the Mined Factor Database. Let $\rho(\hat{f})$ denote the family signature of $\hat{f}$, and let $N_{\mathcal{A}}(\rho(\hat{f}))$ be the number of previously observed factors from the same family in the Mined Factor Database. The family-level shaping term is written as Equation~\ref{eq:family_reward}, where $\mathbb{I}_{\mathrm{new}}(\hat{f})$ indicates that $\hat{f}$ is a high-quality factor from a previously unseen structural family, and $\mathbb{I}_{\mathrm{over}}(\hat{f})$ indicates that $\hat{f}$ belongs to an overused low-quality family. This design allows useful local refinement within a family while discouraging excessive exploitation of weak templates.

\begin{equation}
	r_{\mathrm{fam}}(\hat{f})
	=
	\lambda_{\mathrm{new}} \mathbb{I}_{\mathrm{new}}(\hat{f})
	-
	\lambda_{\mathrm{fam}} \mathbb{I}_{\mathrm{over}}(\hat{f}),
	\label{eq:family_reward}
\end{equation}

For tasks where behavioral profiles are available, DiCo Reward further introduces a complementarity term. Let $b(\hat{f})$ denote the behavior profile of $\hat{f}$, such as prediction vectors in single-asset timing tasks or cross-sectional rank vectors in RankIC tasks. We define the maximum behavioral similarity to elite factors in the Mined Factor Database as Equation~\ref{eq:behavior_corr}.

\begin{equation}
	c_{\max}(\hat{f})
	=
	\max_{f \in \mathcal{A}_{\mathrm{elite}}}
	\mathrm{corr}\bigl(b(\hat{f}), b(f)\bigr).
	\label{eq:behavior_corr}
\end{equation}

The complementarity shaping term is then defined as Equation~\ref{eq:complementarity_shaping}, where $\mathbb{I}_{\mathrm{low}}(\hat{f})$ indicates sufficiently low behavioral correlation with elite factors in the Mined Factor Database, and $[x]_+=\max(x,0)$. This term rewards behaviorally distinct candidates and penalizes candidates that are highly correlated with already mined elite factors.

\begin{equation}
	r_{\mathrm{comp}}(\hat{f}) = \lambda_{\mathrm{low}} \mathbb{I}_{\mathrm{low}}(\hat{f}) - \lambda_{\mathrm{corr}} [c_{\max}(\hat{f})-\tau_{\mathrm{corr}}]_+.
	\label{eq:complementarity_shaping}
\end{equation}

The final reward combines the predictive and shaping components can be calculate as Equation~\ref{eq:dico_final}.

\begin{equation}
	r_{\hat{f}}
	=
	\mathrm{clip}
	\left(
	r_{\mathrm{pred}}
	+
	r_{\mathrm{exact}}
	+
	r_{\mathrm{fam}}
	+
	r_{\mathrm{comp}},
	r_{\min},
	r_{\max}
	\right).
	\label{eq:dico_final}
\end{equation}

In practice, the predictive component remains the dominant signal, while the diversity and complementarity terms serve as lightweight regularization. Different tasks may enable different subsets of these components depending on the available behavioral profiles. For example, cross-sectional RankIC tasks can use rank-vector behavior records, while simpler settings may only use exact-repeat and structural-family records.

Overall, DiCo Reward encourages the Miner LLM to discover factors that are not only empirically strong, but also structurally diverse and behaviorally complementary to the existing Mined Factor Database. This helps reduce mode collapse during RFT and improves the quality of the final factor library.

\section{Evalution}

In this section, we evalute \textsc{QuantEvolver} from 5 perspectives. overall evalution, ablation study, hyberparameter sensitivity, training reward analysis, profitability case study.

\subsection{Experimental Setup}

Unless otherwise stated, \textsc{QuantEvolver} is instantiated with the \texttt{Qwen3-14B} base model. For downstream evaluation, all methods are assessed under a unified post-selection and fusion protocol: candidate factors are first evaluated individually, then ranked by validation performance, filtered by decorrelation, and finally fused into a multi-factor signal using equal weighting. By default, we adopt validation-guided decorrelated selection with a correlation threshold of $0.7$ and report results from the resulting fused portfolio. All experiments are conducted on a local compute server equipped with 160 Intel(R) Xeon(R) CPU cores, 1.8\,TiB RAM, and 8 NVIDIA H20 GPUs.

\subsection{Metrics}

We evaluate the discovered alpha factors from both predictive accuracy and ranking quality. Given a factor $f$, let $z_{i,t}$ denote the factor value of asset $i$ at time $t$, and let $r_{i,t+1}$ denote the future return of the same asset in the next period. We use four commonly adopted metrics: directional accuracy (DirAcc), information coefficient (IC), rank information coefficient (RankIC), and information coefficient information ratio (ICIR).

\textbf{Directional Accuracy (DirAcc).}
DirAcc measures whether the factor correctly predicts the direction of future returns. It is defined as the proportion of samples where the sign of the factor signal is consistent with the sign of the future return as Equation~\ref{eq:diracc}, where $\Omega$ denotes the set of valid asset-time samples and $\mathbb{I}[\cdot]$ is the indicator function. A higher DirAcc indicates stronger directional forecasting ability.

\begin{equation}
	\mathrm{DirAcc}(f) =
	\frac{1}{|\Omega|}
	\sum_{(i,t) \in \Omega}
	\mathbb{I}\left[
	\mathrm{sign}(z_{i,t}) = \mathrm{sign}(r_{i,t+1})
	\right],
	\label{eq:diracc}
\end{equation}

\textbf{Information Coefficient (IC).}
IC measures the linear correlation between factor values and future returns. For each time $t$, we compute the cross-sectional Pearson correlation between $\{z_{i,t}\}_{i=1}^{N_t}$ and $\{r_{i,t+1}\}_{i=1}^{N_t}$ as Equation~\ref{eq:ic_t}, where $N_t$ is the number of valid assets at time $t$.

\begin{equation}
	\mathrm{IC}_t(f) =
	\mathrm{corr}_{i}
	\left(
	z_{i,t}, r_{i,t+1}
	\right),
	\label{eq:ic_t}
\end{equation} 

The overall IC is then computed as the average over all evaluation periods as Equation~\ref{eq:ic}.

\begin{equation}
	\mathrm{IC}(f) =
	\frac{1}{T}
	\sum_{t=1}^{T}
	\mathrm{IC}_t(f).
	\label{eq:ic}
\end{equation}

A higher absolute IC indicates stronger linear association between the factor signal and future returns. In our evaluation, we report IC in the direction where larger values indicate better predictive performance.

\textbf{Rank Information Coefficient (RankIC).}
RankIC measures the rank correlation between factor values and future returns. It is computed as the cross-sectional Spearman correlation at each time step as Equation~\ref{eq:rankic_t} and averaged over all evaluation periods as Equation~\ref{eq:rankic}.

\begin{equation}
	\mathrm{RankIC}_t(f) =
	\mathrm{corr}_{i}
	\left(
	\mathrm{rank}(z_{i,t}),
	\mathrm{rank}(r_{i,t+1})
	\right),
	\label{eq:rankic_t}
\end{equation}

\begin{equation}
	\mathrm{RankIC}(f) =
	\frac{1}{T}
	\sum_{t=1}^{T}
	\mathrm{RankIC}_t(f).
	\label{eq:rankic}
\end{equation}

RankIC is widely used for cross-sectional factor evaluation because it focuses on whether the factor can correctly rank assets by future performance.

\textbf{Information Coefficient Information Ratio (ICIR).}
ICIR measures the stability of IC over time. It is defined as the mean IC divided by the standard deviation of IC as Equation~\ref{eq:icir}, where $\epsilon$ is a small constant for numerical stability. A higher ICIR indicates that the factor not only has strong predictive correlation, but also maintains stable performance across different periods.

\begin{equation}
	\mathrm{ICIR}(f) =
	\frac{
		\mathrm{mean}\left(\{\mathrm{IC}_t(f)\}_{t=1}^{T}\right)
	}{
		\mathrm{std}\left(\{\mathrm{IC}_t(f)\}_{t=1}^{T}\right) + \epsilon
	},
	\label{eq:icir}
\end{equation}

\subsection{Compared Approaches \& Benchmark}

We compare \textsc{QuantEvolver} with four representative LLM-based alpha factor discovery methods. For a fair comparison, all methods use Qwen-3.6-Plus as the backbone LLM and are evaluated under the same factor expression space, market variables, operator definitions, and validation protocol.

\textbf{AlphaBench}~\cite{luoalphabench} is a systematic benchmark framework for evaluating LLMs in formulaic alpha factor mining. It studies LLM capabilities across factor generation, factor evaluation, and factor searching, and serves as a representative prompt-based LLM alpha mining baseline. :contentReference[oaicite:0]{index=0}

\textbf{QuantaAlpha}~\cite{han2026quantaalpha} is an evolutionary LLM-driven alpha mining framework. It treats each end-to-end mining run as a trajectory and improves generated factors through trajectory-level mutation and crossover, while enforcing semantic consistency among hypotheses, factor expressions, and executable code. :contentReference[oaicite:1]{index=1}

\textbf{R\&D-Agent}~\cite{li2026r} is an LLM-based automated research-and-development framework for data-driven solution building. It adopts a dual-agent design, where a Researcher proposes ideas based on performance feedback and a Developer implements or refines solutions based on execution feedback. We adapt it to alpha factor discovery by treating factor proposal, implementation, and empirical evaluation as an iterative data-driven research workflow. :contentReference[oaicite:2]{index=2}

\textbf{Alpha-Jungle}~\cite{shi2026navigating} refers to the LLM-powered MCTS framework for formulaic alpha factor mining. It combines LLM-based symbolic formula generation with Monte Carlo Tree Search, using quantitative backtesting feedback to guide exploration and frequent-subtree avoidance to improve search efficiency. :contentReference[oaicite:3]{index=3}

We evaluate all methods on three benchmarks, denoted as $\mathbf{A}$, $\mathbf{B}$, and $\mathbf{\Gamma}$, which cover single-asset direction prediction, high-frequency cross-sectional factor mining, and daily equity ETF factor discovery.

\textbf{Benchmark $\mathbf{A}$: real-asset 5-minute directional prediction.}
Benchmark $\mathbf{A}$ is constructed from the 5-minute market data of a real liquid asset. At each timestamp, the task is to predict whether the asset price will increase in the next 5-minute interval compared with the current price. Any positive future price change is labeled as an upward movement, while non-positive movement is labeled as non-upward. Since the task is a single-asset binary directional prediction problem, directional accuracy (DirAcc) is used as the primary metric.

\textbf{Benchmark $\mathbf{B}$: high-frequency cross-sectional factor discovery with hourly rebalancing.}
Benchmark $\mathbf{B}$ is constructed from high-frequency multi-asset data collected from a major digital-asset exchange. At each hourly rebalancing timestamp, a discovered factor assigns scores to multiple tradable assets, and the evaluation measures whether these scores can correctly rank assets by their subsequent returns over the next holding period. This benchmark reflects a realistic high-frequency cross-sectional factor mining setting with hourly portfolio rebalancing, where the method must discover factors that generalize across assets and remain effective under frequent market updates. We use IC, RankIC, and ICIR as evaluation metrics, with RankIC as the primary metric.

\textbf{Benchmark $\mathbf{\Gamma}$: daily CSI 300 ETF factor discovery.}
Benchmark $\mathbf{\Gamma}$ is constructed from daily market data of CSI 300 ETF constituents. The task is to discover daily alpha factors that predict future asset returns or cross-sectional rankings under a standard equity factor evaluation protocol. Compared with Benchmarks $\mathbf{A}$ and $\mathbf{B}$, this benchmark represents a lower-frequency and more conventional equity-market factor discovery setting. We evaluate generated factors using IC, RankIC, and ICIR, and use RankIC as the primary metric.

Together, the three benchmarks evaluate alpha factor discovery under complementary settings: Benchmark $\mathbf{A}$ focuses on short-horizon single-asset direction prediction, Benchmark $\mathbf{B}$ focuses on high-frequency cross-sectional ranking, and Benchmark $\mathbf{\Gamma}$ focuses on daily equity ETF factor discovery.

\subsection{Overall Evaluation}

\begin{table*}[t]
	\centering
	\small
	\setlength{\tabcolsep}{3.2pt}
	\renewcommand{\arraystretch}{1.18}
	\caption{Overall Evaluation}
	\label{tab:overall_evaluation}
	
	\newcolumntype{H}{>{\columncolor{acccolor}}c}
	
	\begin{tabular}{
			l
			@{\hspace{8pt}}
			H c c c
			@{\hspace{12pt}}
			c c H c
			@{\hspace{12pt}}
			c c H c
		}
		\toprule
		\multirow{2}{*}{\textbf{Method}}
		& \multicolumn{4}{c}{\cellcolor{latcolor}$\mathbf{A}$}
		& \multicolumn{4}{c}{\cellcolor{latcolor}$\mathbf{B}$}
		& \multicolumn{4}{c}{\cellcolor{latcolor}$\mathbf{\Gamma}$} \\
		\cmidrule(lr{6pt}){2-5}
		\cmidrule(lr{6pt}){6-9}
		\cmidrule(lr{6pt}){10-13}
		& \textbf{DirAcc*}
		& \textbf{IC}
		& \textbf{RankIC}
		& \textbf{ICIR}
		& \textbf{DirAcc}
		& \textbf{IC}
		& \textbf{RankIC*}
		& \textbf{ICIR}
		& \textbf{DirAcc}
		& \textbf{IC}
		& \textbf{RankIC*}
		& \textbf{ICIR} \\
		\midrule
		AlphaBench
		& 51.82\%
		& \textbf{0.0147}
		& 0.0362
		& 0.6139
		& \underline{49.57\%}
		& \underline{0.0288}
		& \underline{0.0337}
		& 24.4040
		& \textbf{55.56\%}
		& \underline{0.1480}
		& \underline{0.1688}
		& \underline{2.6021} \\
		
		QuantaAlpha
		& 52.15\%
		& \underline{0.0095}
		& 0.0313
		& \textbf{3.6004}
		& 49.41\%
		& 0.0216
		& 0.0235
		& \underline{24.4209}
		& \underline{54.03\%}
		& 0.1144
		& 0.1003
		& 2.0239 \\
		
		R\&D-Agent
		& \underline{52.59\%}
		& 0.0002
		& \textbf{0.0527}
		& 0.9779
		& 49.40\%
		& 0.0188
		& 0.0205
		& 19.0101
		& 53.31\%
		& 0.1001
		& 0.0589
		& -0.8667 \\
		
		Alpha-Jungle
		& 52.25\%
		& 0.0023
		& 0.0357
		& \underline{1.5664}
		& 49.39\%
		& 0.0185
		& 0.0200
		& 19.2806
		& 48.92\%
		& 0.0369
		& -0.0008
		& -0.7166 \\
		
		\textbf{\textsc{QuantEvolver}}
		& \textbf{53.22\%}
		& -0.0058
		& \underline{0.0481}
		& 1.1191
		& \textbf{49.96\%}
		& \textbf{0.0500}
		& \textbf{0.0586}
		& \textbf{50.2644}
		& 53.49\%
		& \textbf{0.1502}
		& \textbf{0.1923}
		& \textbf{5.4289} \\
		\bottomrule
	\end{tabular}
\end{table*}

Table~\ref{tab:overall_evaluation} reports the overall performance of \textsc{QuantEvolver} and representative LLM-based alpha factor discovery baselines on the three benchmarks. The highlighted columns indicate the primary metric of each benchmark: DirAcc for Benchmark~$\mathbf{A}$ and RankIC for Benchmarks~$\mathbf{B}$ and $\mathbf{\Gamma}$. Overall, \textsc{QuantEvolver} achieves the best performance on the primary metric across all three benchmarks, demonstrating its effectiveness under heterogeneous factor discovery settings.

On Benchmark~$\mathbf{A}$, which focuses on short-horizon single-asset directional prediction, \textsc{QuantEvolver} achieves the highest DirAcc of 53.22\%, outperforming the strongest baseline R\&D-Agent at 52.59\%. Although several baselines obtain higher auxiliary IC or ICIR values, these metrics are not the primary objective of this directional prediction task. The improvement in DirAcc indicates that reinforcement fine-tuning helps the Miner LLM discover factors that are better aligned with the target decision rule of next-interval price direction.

On Benchmark~$\mathbf{B}$, which evaluates high-frequency cross-sectional factor mining, \textsc{QuantEvolver} obtains the strongest RankIC of 0.0586, substantially exceeding the best baseline RankIC of 0.0337. It also achieves the highest IC and ICIR, reaching 0.0500 and 50.2644, respectively. These results show that \textsc{QuantEvolver} not only improves the primary ranking metric, but also produces factors with stronger linear correlation and more stable information coefficients over time. The consistent advantage across IC, RankIC, and ICIR suggests that the discovered factors generalize better across assets in a high-frequency cross-sectional setting.

On Benchmark~$\mathbf{\Gamma}$, the daily equity ETF factor discovery benchmark, \textsc{QuantEvolver} achieves the best RankIC of 0.1923 and the best ICIR of 5.4289. It also obtains the highest IC of 0.1502, slightly outperforming AlphaBench. Although AlphaBench achieves the highest DirAcc, RankIC is the primary metric for this benchmark because the task focuses on factor ranking quality rather than binary direction prediction. The strong RankIC and ICIR results indicate that \textsc{QuantEvolver} can discover daily equity factors with both high predictive ranking quality and stable out-of-sample behavior.

These results support two main observations. First, \textsc{QuantEvolver} consistently improves the primary task metric across all benchmarks, showing that policy-update-driven factor discovery is effective beyond a single market or evaluation protocol. Second, the gains are most pronounced on the two ranking-oriented benchmarks, where the ability to generate diverse and complementary factor candidates is especially important. This confirms that \textsc{QuantEvolver} is not merely optimizing isolated factor expressions, but learning a stronger factor generation policy that transfers across different alpha discovery scenarios.

\subsection{Ablation Study}

Table~\ref{tab:ablation_core} presents the ablation results on Benchmark~$\mathbf{B}$, where RankIC is the primary metric. We study three key components of \textsc{QuantEvolver}: seed construction (\textbf{Seed}), diversity-aware reward shaping (\textbf{Div}), and the Factor DSL constraint (\textbf{DSL}). The full model, which uses all three components, achieves the best performance across all metrics, with an IC of 0.0500, a RankIC of 0.0586, and an ICIR of 50.2644.

\begin{table}[htbp]
	\centering
	\small
	\setlength{\tabcolsep}{9.6pt}
	\renewcommand{\arraystretch}{1.15}
	\caption{Ablation Results on Dataset $\mathbf{B}$}
	\label{tab:ablation_core}
	\begin{tabular}{ccc|c>{\columncolor{acccolor}}c c}
		\toprule
		\textbf{Seed} & \textbf{Div} & \textbf{DSL}
		& \textbf{IC}
		& \textbf{RankIC*}
		& \textbf{ICIR} \\
		\midrule
		\checkmark & \checkmark & \checkmark & \textbf{0.0500} & \textbf{0.0586} & \textbf{50.2644} \\
		\checkmark &            & \checkmark & 0.0421          & 0.0505          & 45.8038 \\
		            & \checkmark & \checkmark & 0.0434          & 0.0519          & 44.2189 \\
		            &            & \checkmark & 0.0461          & 0.0540          & 44.3755 \\
		            &            &            & 0.0001          & 0.0002          & 0.2506 \\
		\bottomrule
	\end{tabular}
\end{table}

Removing either seed construction or diversity shaping leads to a clear performance drop. Without diversity shaping, RankIC decreases from 0.0586 to 0.0505, suggesting that the Miner LLM tends to over-exploit similar factor patterns when diversity and complementarity signals are absent. Without seed construction, RankIC decreases to 0.0519, indicating that high-quality seed factors provide useful starting points for RFT and help guide the search toward promising regions of the factor space. These results show that both seed initialization and diversity-aware reward shaping contribute to effective factor discovery.

Interestingly, the variant without both Seed and Div but still using the Factor DSL achieves a RankIC of 0.0540, which is lower than the full model but remains competitive. This suggests that the Factor DSL itself provides an important inductive bias by restricting the search space to executable and financially meaningful factor expressions. However, adding seed construction and diversity-aware shaping on top of the DSL further improves both factor quality and stability.

The last row removes the Factor DSL and lets the model generate unconstrained factor code. Its performance collapses to a RankIC of only 0.0002 and an ICIR of 0.2506. This demonstrates that unconstrained code generation is unreliable for RFT-based factor discovery: generated programs are more likely to be invalid, unstable, or poorly aligned with the evaluation objective. In contrast, the Factor DSL provides a structured action space that makes reinforcement fine-tuning feasible and sample-efficient.

Overall, the ablation study confirms that the strongest performance comes from the combination of all three components. The Factor DSL ensures executable and well-formed factor generation, seed construction provides high-quality starting points, and diversity-aware reward shaping prevents repetitive exploitation while encouraging complementary discoveries.

\subsection{Hyperparameter Sensitivity}

\begin{figure}[htbp]
	\centering
	\subfigure[Top-$k$ Sensitivity]{
		\begin{minipage}{0.46\linewidth}
			\centering
			\includegraphics[width=\textwidth]{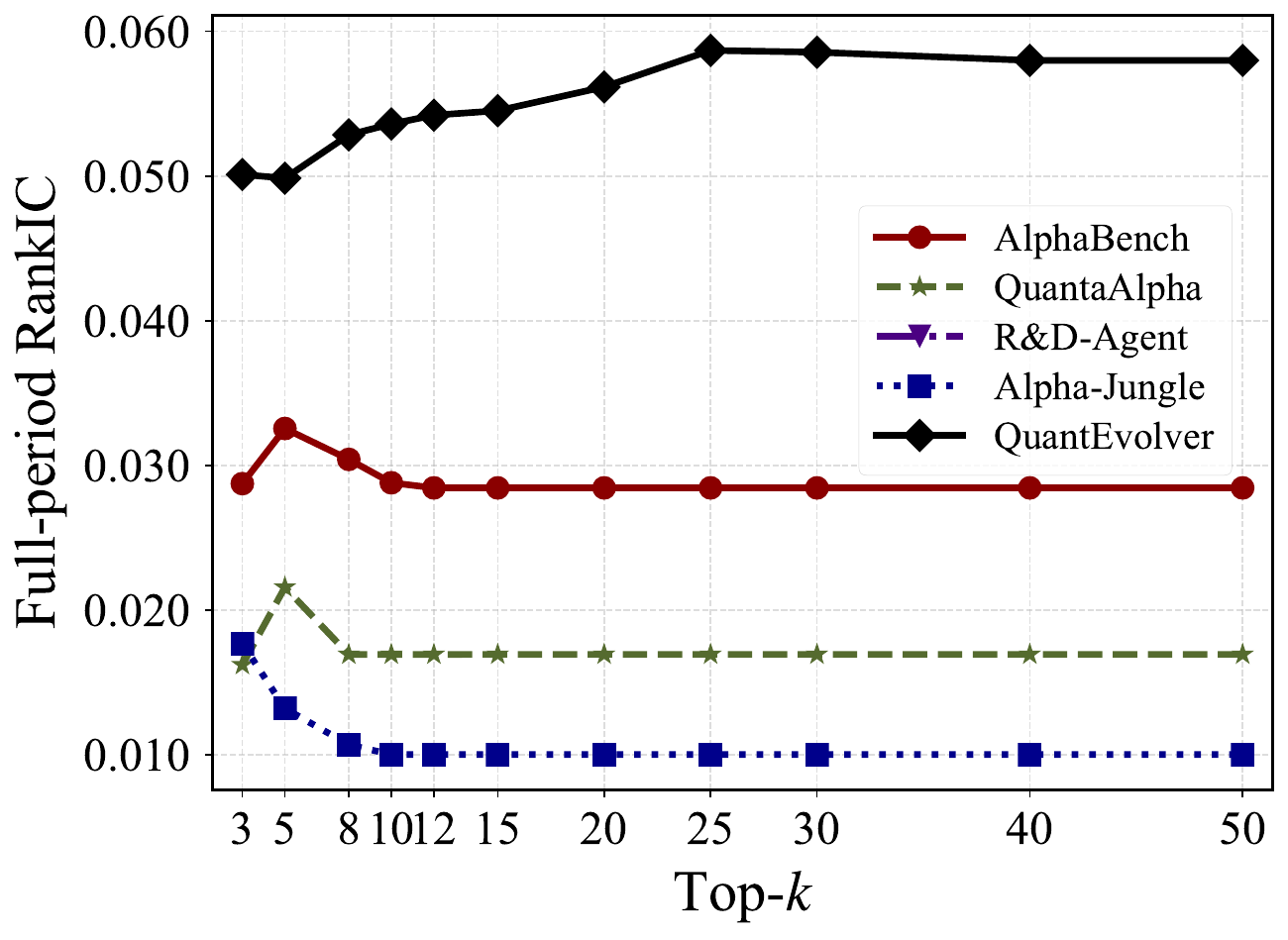}
			\label{fig:topk_sensitivity_xsec}
		\end{minipage}
	}
	\subfigure[Correlation-Threshold Sensitivity]{
		\begin{minipage}{0.46\linewidth}
			\centering   
			\includegraphics[width=\textwidth]{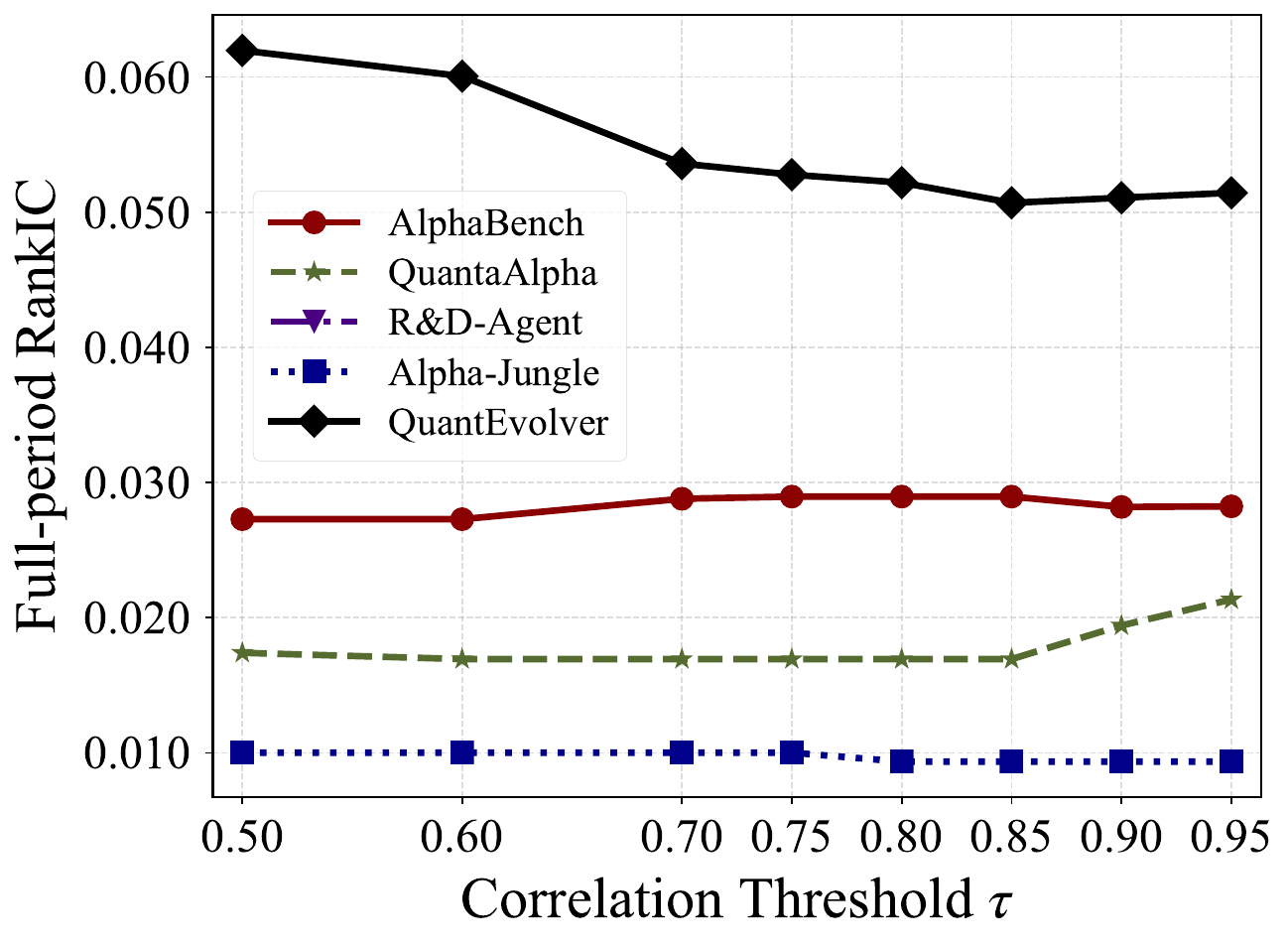}
			\label{fig:corrth_sensitivity_xsec}
		\end{minipage}
	}
	\caption{Hyberparameter Sensitivity Experiment on Dataset $\mathbf{B}$}
	\label{fig:hyberparameter-sensitivity}
\end{figure}

Figure~\ref{fig:hyberparameter-sensitivity} studies the sensitivity of factor fusion performance to two important hyperparameters on Benchmark~$\mathbf{B}$: the number of selected factors $k$ and the correlation threshold used during factor selection. Since Benchmark~$\mathbf{B}$ uses RankIC as the primary metric, we report the fused out-of-sample RankIC under different hyperparameter settings.

Figure~\ref{fig:topk_sensitivity_xsec} shows that \textsc{QuantEvolver} consistently outperforms all baselines across different top-$k$ settings. Its RankIC remains around 0.050 even when only the top three to five factors are selected, and steadily improves as more high-quality factors are included. The performance increases from 0.0498 at $k=5$ to 0.053 at $k=10$, 0.0562 at $k=20$, and reaches its best value of 0.0587 around $k=25$. The result remains nearly unchanged at $k=30$ with a RankIC of 0.0586, and only slightly decreases when $k$ is further enlarged. This trend indicates that \textsc{QuantEvolver} produces a rich and stable factor pool: adding more selected factors improves the fused signal up to a moderate scale, while the performance remains robust when the selected set becomes larger.

Figure~\ref{fig:corrth_sensitivity_xsec} examines the effect of the correlation threshold used during factor selection. A lower threshold imposes a stricter decorrelation constraint among selected factors. \textsc{QuantEvolver} achieves the strongest performance under stricter thresholds, reaching a RankIC of 0.0620 at threshold 0.50 and 0.0601 at threshold 0.60. As the threshold becomes looser, the RankIC decreases to around 0.052--0.054, suggesting that controlling redundancy among selected factors is important for obtaining a stronger fused signal. This result indicates that the mined factor pool contains many individually predictive candidates, but effective fusion still benefits from explicitly preserving complementary information.

Overall, the sensitivity analysis shows that \textsc{QuantEvolver} is robust to fusion hyperparameters. It maintains a clear advantage over baseline methods across a wide range of top-$k$ values and correlation thresholds, while the best configurations favor a moderately sized factor set with relatively strict decorrelation. This further supports that \textsc{QuantEvolver} discovers not only isolated strong factors, but also a diverse and complementary factor pool suitable for multi-factor fusion.

\subsection{Mining Process Analysis}

\begin{figure}[htbp]
	\centering
	\subfigure[Training Reward Dynamics]{
		\begin{minipage}{0.46\linewidth}
			\centering
			\includegraphics[width=\textwidth]{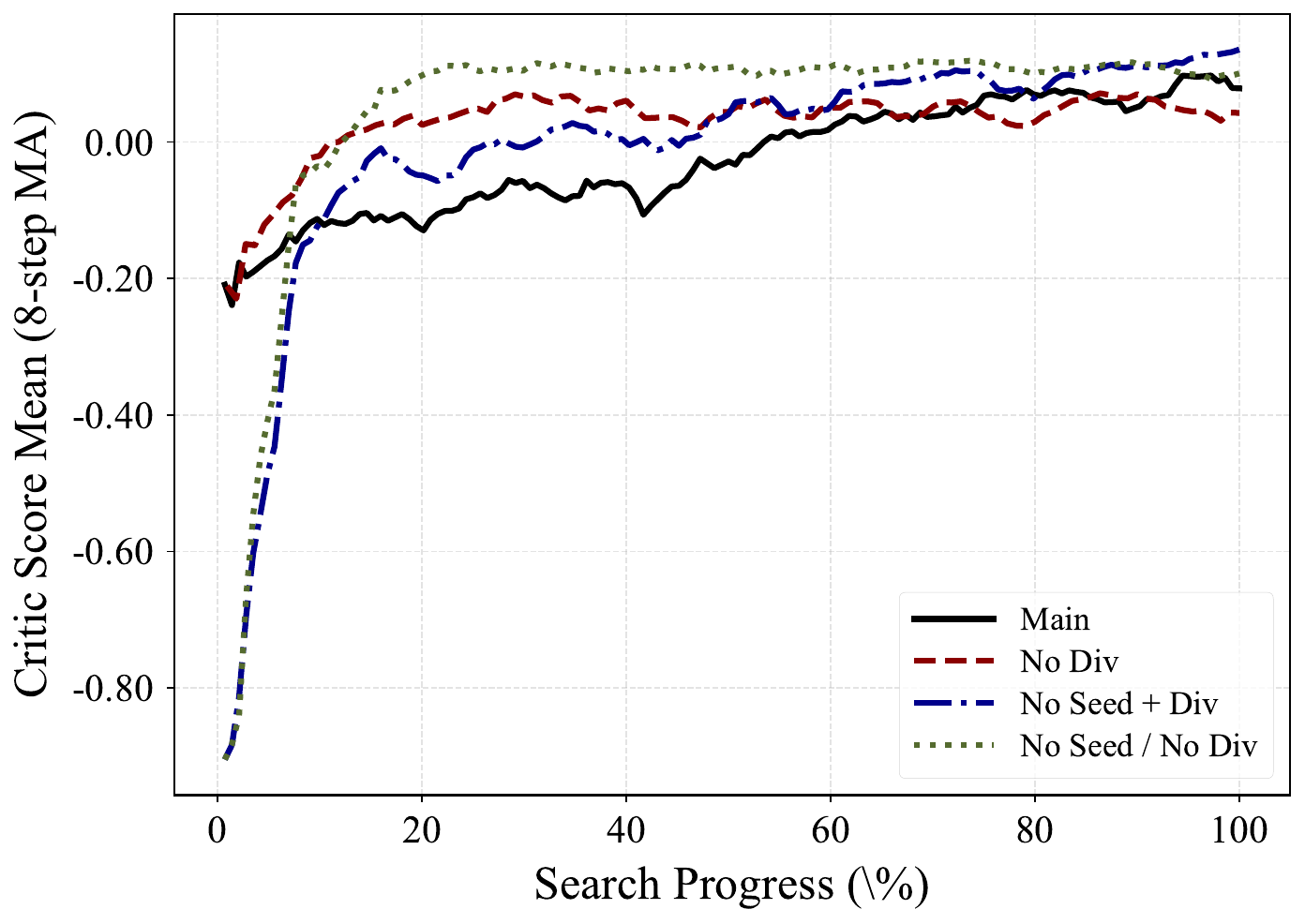}
			\label{fig:critic_score_mean}
		\end{minipage}
	}
	\subfigure[High-Quality Factor Discovery]{
		\begin{minipage}{0.46\linewidth}
			\centering   
			\includegraphics[width=\textwidth]{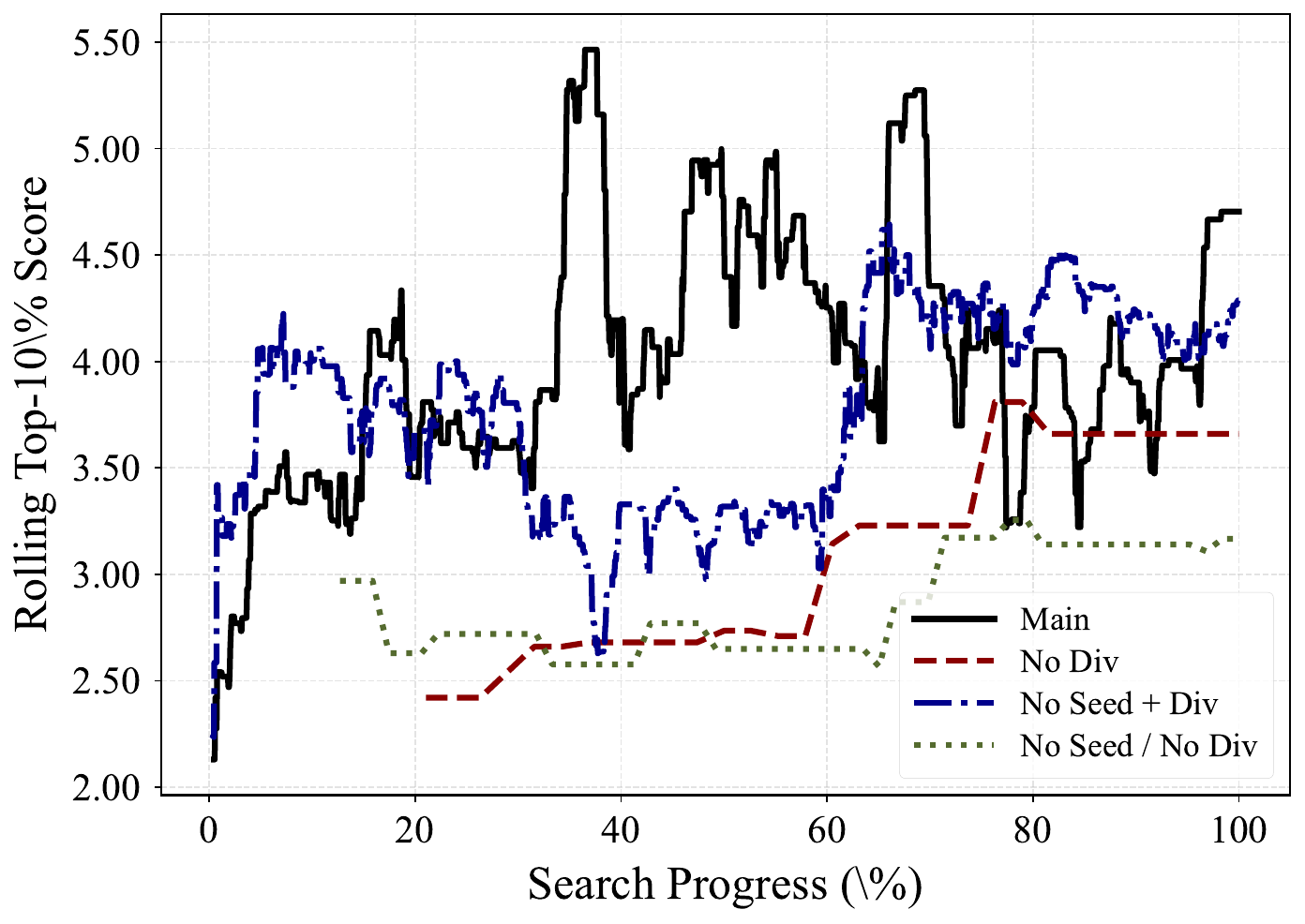}
			\label{fig:rolling_top10_score}
		\end{minipage}
	}
	\caption{Mining process analysis on Dataset $\mathbf{B}$.}
	\label{fig:mining_process_analysis}
\end{figure}

Figure~\ref{fig:mining_process_analysis} provides a closer look at the mining dynamics of \textsc{QuantEvolver} and its ablated variants on Dataset~$\mathbf{B}$. In Figure~\ref{fig:critic_score_mean}, the full \textsc{QuantEvolver} design exhibits a more stable and consistently improving reward trajectory over search progress, indicating that the combination of seed initialization and diversity-aware exploration helps the search process move toward more promising regions of the factor space. Removing either component leads to a weaker or less stable optimization trend, while removing both causes the mining process to become substantially less effective.

Figure~\ref{fig:rolling_top10_score} further examines the quality of high-scoring discovered factors. The full method maintains the strongest rolling top-10\% score over most of the search process, showing that \textsc{QuantEvolver} not only improves the average optimization trajectory but also discovers stronger high-quality candidates more reliably. In contrast, the ablated variants either improve more slowly or plateau at lower quality levels. These results suggest that seed-based initialization and diversity-aware search are both important for sustaining effective factor discovery throughout the mining process.

\subsection{Profitability Case Study}

\begin{figure}[htbp]
	\centering
	\includegraphics[width=1\linewidth]{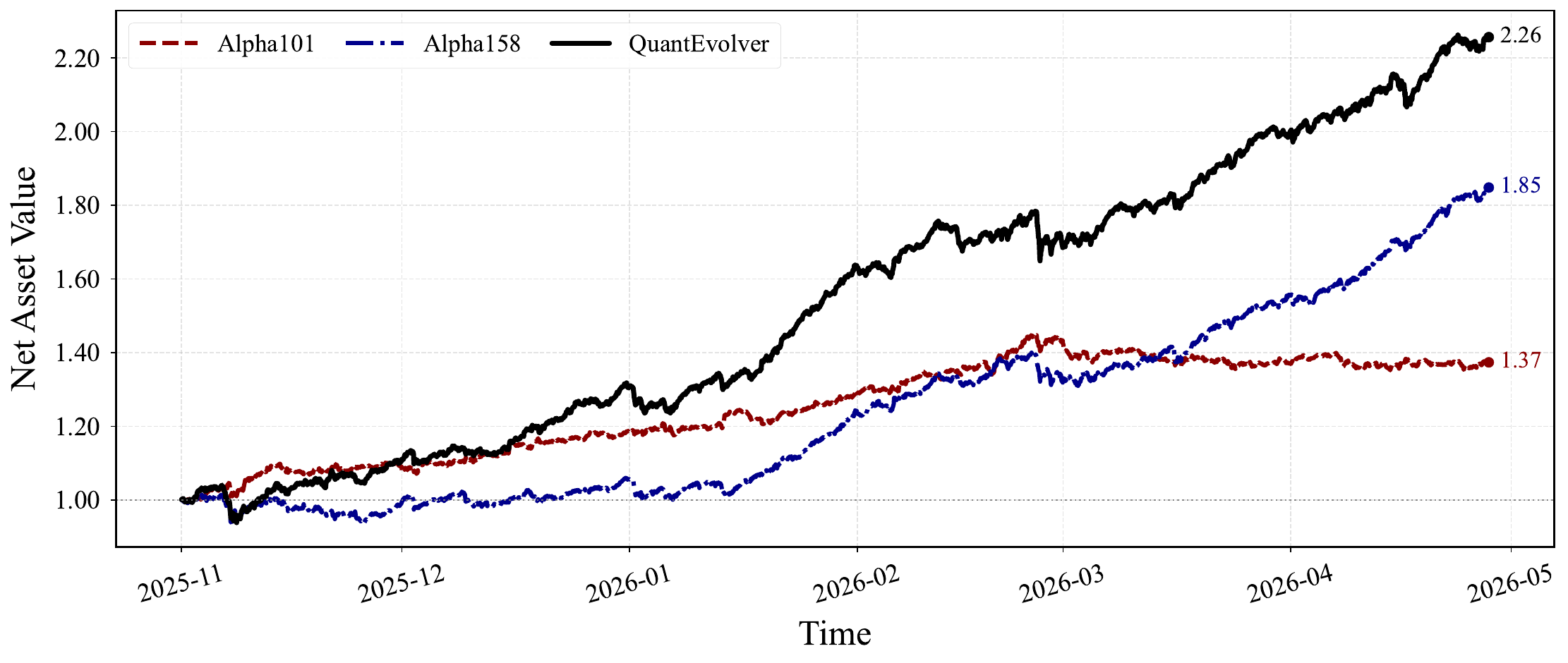}
	\caption{Cumulative Return Comparison on Dataset $\mathbf{B}$}
	\label{fig:profitability}
\end{figure}

Figure~\ref{fig:profitability} presents a profitability case study on Benchmark~$\mathbf{B}$ by converting discovered cross-sectional factors into a simple long--short portfolio. At each rebalancing timestamp, assets are ranked by the factor signal, and the portfolio takes long positions in the top-ranked assets and short positions in the bottom-ranked assets. This experiment is not intended to be a fully optimized trading system; instead, it serves as a practical sanity check on whether the higher RankIC achieved by \textsc{QuantEvolver} can translate into economically meaningful returns.

The results show that the \textsc{QuantEvolver}-based portfolio produces the strongest cumulative return among all compared factor sets. Over the evaluation period from November 2025 to April 2026, the net asset value of \textsc{QuantEvolver} increases from 1.00 to about $2.26$, corresponding to a cumulative return of approximately $125.6\%$. In comparison, Alpha158 reaches a final net asset value of about $1.85$, while Alpha101 reaches about $1.37$. Thus, \textsc{QuantEvolver} not only improves the statistical ranking metric reported in previous sections, but also leads to a substantially stronger long--short return curve.

The return trajectory also indicates that \textsc{QuantEvolver} maintains a persistent advantage over the classical alpha sets. Although all methods experience short-term fluctuations and drawdowns, the \textsc{QuantEvolver} curve begins to separate from the baselines after the early evaluation period and continues to compound more rapidly. This suggests that the mined factors capture cross-sectional return signals that remain effective over time rather than producing only isolated high-IC observations.

Overall, this case study provides additional evidence that the factor pool discovered by \textsc{QuantEvolver} is not merely statistically predictive, but also economically useful under a simple portfolio construction protocol. The stronger cumulative return supports the practical relevance of policy-update-driven factor discovery for high-frequency cross-sectional alpha mining.

\section{Related Work}

\subsection{LLM-Based Quantitative Finance}

With the rapid development of large language models (LLMs), increasing efforts have explored their applications in quantitative finance. From the perspective of financial tasks, existing work can be broadly divided into five categories: data analysis, investment research, trading, investment management, and risk management~\cite{dong2025large}.

Data analysis focuses on using LLMs to process heterogeneous financial information, such as earnings-call transcripts, annual reports, financial news, SEC filings, and social media posts. Representative tasks include financial text summarization, named entity recognition, relation extraction, event classification, and sentiment analysis~\cite{mukherjee2022ectsum, li2023lcfn, sharma2022finred}. Domain-specific financial LLMs, such as FinBERT, BloombergGPT, FinGPT, FinMA, InvestLM, FinTral, and ICE-Intern, further improve financial language understanding by adapting general language models to financial corpora or instruction data~\cite{liu2021finbert, wu2023bloomberggpt, yang2023fingpt, xie2023pixiu, yang2023investlm, bhatia2024fintral, hu2024no}.

Investment research aims to support market analysis, asset evaluation, forecasting, and investment hypothesis generation. Existing work studies LLMs for event classification, financial sentiment analysis, time-series forecasting, and investment reasoning over market data and textual information~\cite{lee2021fednlp, shah2023trillion, sinha2021impact, yu2023harnessing, hu2025fintsb}. More closely related to this paper, recent studies apply LLMs to alpha factor mining, where LLMs are used as factor miners, evaluators, interactive assistants, or search controllers to generate and refine formulaic alpha expressions~\cite{wang2025alpha, li2024can, wang2024quantagent, luoalphabench, tang2025alphaagent, shi2026navigating, li2026r, han2026quantaalpha}.

Trading focuses on using LLMs to support strategy execution and trading decision-making. In this setting, LLMs may interpret market states, summarize relevant signals, reason over trading opportunities, or generate executable trading decisions based on historical prices, news, reports, and sentiment information~\cite{gupta2023gpt, xie2024finben, li2025investorbench, lu2025strux}. Other works further explore memory-augmented or agentic trading systems, where LLM agents coordinate market observation, reflection, and decision-making over time~\cite{yu2025finmem, li2023cfgpt, zhang2024ai, xiao2025tradingagents}.

Investment management studies how LLMs can assist portfolio-level reasoning, asset allocation, financial question answering, and investment recommendation. Representative benchmarks and models focus on answering complex financial questions over reports, tables, and market information, requiring numerical reasoning, multi-hop reasoning, and risk-return trade-off analysis~\cite{chen2022convfinqa, choi2025finder, li2024alphafin}. These studies show the potential of LLMs to support portfolio construction and investment decision workflows, but also reveal challenges in long-context reasoning, numerical calculation, and adaptation to dynamic financial data.

Risk management applies LLMs to tasks such as fraud detection, default risk prediction, compliance analysis, and financial risk monitoring. Existing studies and benchmarks examine how LLMs can identify abnormal transactions, assess borrower risk, and support regulatory or compliance-oriented analysis~\cite{balasubramanian2022substituting, yin2023finpt, feng2023empowering}. These applications require models to handle imbalanced data, evolving risk patterns, and interpretable financial evidence.

Overall, LLM-based quantitative finance has evolved from financial information processing toward investment research, trading support, portfolio reasoning, and risk monitoring. This paper belongs to the investment research direction, with a specific focus on LLM-based alpha factor discovery.

\subsection{Alpha Factor Discovery}

Alpha factor discovery aims to automatically identify predictive signals from historical market data. Existing methods can be broadly categorized according to how they explore the factor expression space.

Early automated methods mainly rely on rule-based symbolic search. These methods represent alpha factors as symbolic expressions and search over the expression space using genetic programming, symbolic regression, evolutionary search, operator-template enumeration, or grammar-based constraints~\cite{zhang2020autoalpha, ren2024alpha, kakushadze2016101, arnaldo2014multiple, yang2026alpha, ren2024riskminer}. By explicitly defining variables, operators, and composition rules, symbolic-search methods can generate executable and interpretable factor expressions. However, their exploration process is often constrained by hand-crafted mutation rules, crossover strategies, search heuristics, or syntactic templates, making it difficult to scale efficiently to large and semantically rich factor spaces.

Another line of work introduces learning-guided factor generation. Instead of relying purely on heuristic symbolic search, these methods use trainable models, reinforcement learning policies, or neural generators to guide the construction, ranking, and selection of alpha expressions~\cite{yu2023generating, xu2024text, zhu2025alphaqcm, shi2025alphaforge, chen2025alphasage, yang2026alpha}. For example, reinforcement-learning-based methods formulate factor discovery as a sequential expression-construction problem, where an agent selects variables and operators according to empirical rewards. Neural and generative methods further learn factor-generation patterns from historical evaluations or previously discovered factors. These approaches improve search adaptivity, but they usually require predefined action spaces, expression templates, or task-specific training objectives, and their generation ability is still limited by the learned search policy rather than open-ended semantic composition.

Recently, LLM-based methods have provided a new paradigm for alpha factor discovery. Leveraging their symbolic reasoning, code-like generation, and instruction-following capabilities, LLMs can generate and refine factor expressions from alpha libraries, market contexts, user instructions, and empirical feedback~\cite{wang2025alpha, li2024can, wang2024quantagent, luoalphabench, tang2025alphaagent, shi2026navigating, li2026r, han2026quantaalpha}. Existing LLM-based frameworks have explored interactive factor mining, factor evaluation and refinement, search-augmented generation, multi-agent workflows, and evolutionary exploration. These methods improve the flexibility of alpha mining, but most of them still use LLMs as prompt-driven miners, evaluators, or agents.

Our work is positioned at the intersection of learning-guided factor generation and LLM-based alpha mining. Unlike rule-based symbolic search, \textsc{QuantEvolver} leverages a language-model policy to generate DSL factor expressions. Unlike existing prompt-driven LLM-based methods, it uses reinforcement fine-tuning as a search-distribution optimization mechanism during factor mining. The goal is not merely to obtain a deployable factor-mining agent, but to collect, validate, and select a high-quality mined factor set throughout the training process.

\section{Conclusion}

In this paper, we study LLM-based alpha factor discovery and identify two key limitations of existing prompt-level optimization methods: context explosion and search stagnation. To address these challenges, we propose \textsc{QuantEvolver}, a self-evolving alpha factor discovery framework based on reinforcement fine-tuning. Instead of accumulating historical candidates and feedback in the prompt, \textsc{QuantEvolver} converts executable quantitative evaluation into policy updates, enabling a Miner LLM to internalize optimization experience through parameter learning and progressively improve its factor generation policy. Experiments on three realistic market benchmarks demonstrate that \textsc{QuantEvolver} consistently discovers stronger and more robust alpha factors than representative LLM-based baselines.

\balance
\bibliographystyle{IEEEtran}
\bibliography{mylib}

\end{document}